\documentclass[runningheads]{llncs}
\usepackage[T1]{fontenc}
\usepackage{graphicx}
\usepackage{amsmath}
\usepackage{algorithm}
\usepackage{amssymb}
\usepackage{makecell}
\usepackage{multicol,multirow,booktabs}
\usepackage[noend]{algpseudocode}
\usepackage[
    colorlinks=true,
    linkcolor=blue,
    citecolor=blue,
    urlcolor=blue
]{hyperref}
\usepackage[misc]{ifsym}
\begin{document}
\title{MORE-R1: Guiding LVLM for Multimodal Object-Entity Relation Extraction via Stepwise Reasoning with Reinforcement Learning}
\titlerunning{MORE-R1}
% If the paper title is too long for the running head, you can set
% an abbreviated paper title here
%
\author{
Xiang Yuan\inst{1}
\and Xu Chu\inst{1}
\and Xinrong Chen\inst{1}
\and Haochen Li\inst{2}
\and Zonghong Dai\inst{2}
\and Hongcheng Fan\inst{2}
\and Xiaoyue Yuan\inst{3}
\and Weiping Li\inst{1}
\and Tong Mo\inst{1}\textsuperscript{(\Letter)}
}
\authorrunning{X. Yuan et al.}
% First names are abbreviated in the running head.
% If there are more than two authors, 'et al.' is used.
%
\institute{
School of Software and Microelectronics, Peking University, Beijing, China
\email{xiangyuan@stu.pku.edu.cn, motong@ss.pku.edu.cn}
\and AlignBase, Beijing, China
\and Information Application Research Center, Shanghai Municipal Administration\\ 
for Market Regulation, Shanghai, China}
\maketitle
\setcounter{footnote}{0}
\vspace{-1em}
% typeset the header of the contribution
%
\begin{abstract}
Multimodal Object-Entity Relation Extraction (MORE) is a challenging task in information extraction research. It aims to identify relations between visual objects and textual entities, requiring complex multimodal understanding and cross-modal reasoning abilities. Existing methods, mainly classification-based or generation-based without reasoning, struggle to handle complex extraction scenarios in the MORE task and suffer from limited scalability and intermediate reasoning transparency. To address these challenges, we propose MORE-R1, a novel model that introduces explicit stepwise reasoning with Reinforcement Learning (RL) to enable Large Vision-Language Model (LVLM) to address the MORE task effectively. MORE-R1 integrates a two-stage training process, including an initial cold-start training stage with Supervised Fine-Tuning (SFT) and a subsequent RL stage for reasoning ability optimization. In the initial stage, we design an efficient way to automatically construct a high-quality SFT dataset containing fine-grained stepwise reasoning tailored to the MORE task, enabling the model to learn an effective reasoning paradigm. In the subsequent stage, we employ the Group Relative Policy Optimization (GRPO) RL algorithm with a Progressive Sample-Mixing Strategy to stabilize training and further enhance model's reasoning ability on hard samples. Comprehensive experiments on the MORE benchmark demonstrate that MORE-R1 achieves state-of-the-art performance with significant improvement over baselines.

\end{abstract}
\section{Introduction} \label{sec1}
\begin{figure}[t]
\centering
\includegraphics[width=0.98\textwidth]{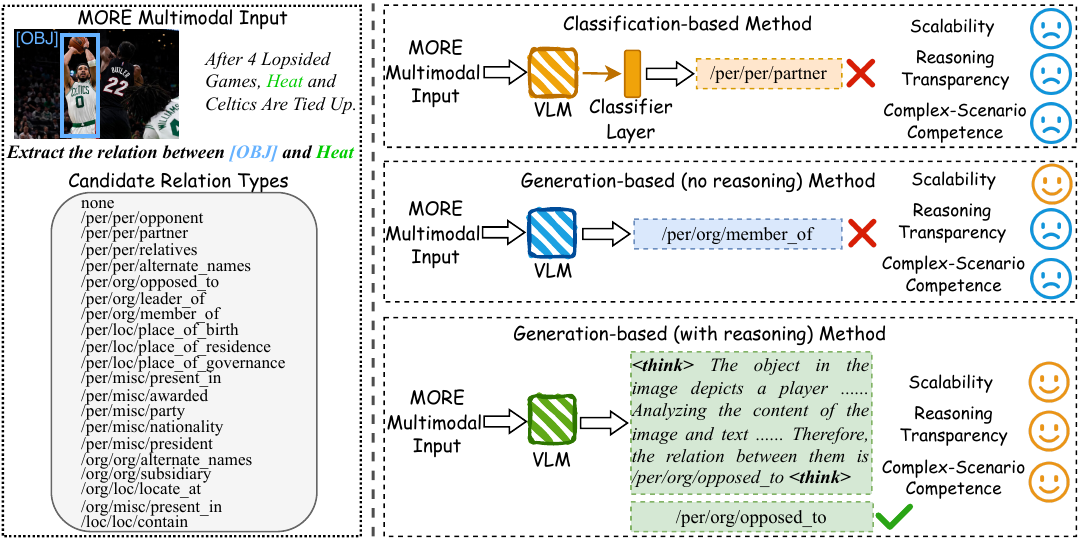}
\caption{Left side: An example of the MORE task: Identify the relation type between the object and entity from the candidate set. Right side: Different frameworks for the MORE task. Our model belongs to the Generation-based (with reasoning) Method.} \label{fig1}
\vspace{-1em}
\end{figure}

Relation Extraction (RE) is a fundamental task in Information Extraction (IE) research, which aims to extract relations between entities from unstructured inputs. Among various RE-related tasks, Multimodal Object-Entity Relation Extraction (MORE) \cite{moreformer} is a newly proposed RE task that better reflects real-world scenarios, aligning with the growing need to process multimodal information (primarily image and text) on the Internet \cite{zhao2024comprehensive}. The MORE task aims to extract the relation between the given object in the image and the given entity in the text, requiring models to pay equal attention to both modalities, establish correspondences between objects and entities, and perform cross-modal relation reasoning. As illustrated by an example in Fig. \ref{fig1}, to precisely extract the relation between the object enclosed by the blue bounding box in the image and the entity ``Heat'' in the text, the model needs to comprehend that the object represents a basketball player from the ``Celtics'' team mentioned in text and the entity ``Heat'' in text is a basketball team name and corresponds to a team member wearing a black jersey with the number ``22'' shown in the image. The model needs to infer that these two teams are competing against each other to accurately predict the ``opposed to'' relation. Therefore, the MORE task imposes higher demands on the model's ability for complex multimodal reasoning. It provides crucial support for various downstream applications such as cross-modal retrieval \cite{liu2024semscene} and multimodal knowledge graph construction \cite{chen2024continual}.

Most existing MORE works \cite{moreformer,mmib,cgimre,focalmre,hvformer,remote} are classification-based: they build upon small-scale pretrained Vision-Language Models (VLMs) using a BERT-based text encoder and a ViT-based visual encoder. These works primarily focus on enhancing fine-grained cross-modal representations of the object and entity, then use a classifier layer to map the features into a set of predefined discrete relation categories. Despite the progress they have made, Classification-based Method falls short in two aspects. (1) Poor scalability. This type of method can only recognize relations belonging to the predefined category set. When new relation types arrive, the model needs to redesign the classifier layer and retrain its network parameters. (2) Unable to handle complex RE scenario in the MORE task. Classification method has limited semantic expressiveness and discriminability for different relation category labels, failing to effectively understand and distinguish between easily confusable relations, such as ``peer'' and ``couple''.
 
Recently, large-scale pretrained VLMs have evolved rapidly. Among them, Large Vision-Language Models (LVLMs) that integrate vision encoders with Large Language Models (LLMs), such as LLaMA3.2-Vision \cite{llama3.2vision} and Qwen2.5-VL \cite{qwen2.5vl}, have achieved impressive results in diverse multimodal understanding and reasoning tasks. To adapt LVLMs to the MORE task, a straightforward approach is to design task-specific prompts or apply instruction tuning that directly forces the model to output relation labels, which falls under the Generation-based (no reasoning) Method in Fig. \ref{fig1}. Although it ensures a certain degree of scalability, recent studies \cite{xing2025benchmarking,yang2023mm} have shown that LLMs and LVLMs with prompt learning or direct fine-tuning on RE labels perform poorly on various RE tasks. These works attribute this limitation to the scarcity of RE-related data during LLMs' and LVLMs' pre-training and instruction tuning stage, as well as the high semantic complexity of discrete relation labels, which hinders models from effectively understanding and following RE-specific instructions \cite{chen2023chain,yang2023mm}. Consequently, such method remains inadequate for the complex reasoning required in MORE. Moreover, directly generating relation labels provides no insight into the model's decision process, undermining both interpretability and reasoning transparency.

Recently advanced large reasoning models such as OpenAI-o1 \cite{jaech2024openai} and Deep-
seek-R1 \cite{guo2025deepseek} have demonstrated remarkable capabilities across a wide range of challenging tasks such as programming and mathematics. These models achieve this by incorporating extended Chain-of-Thought (CoT) reasoning and employing Reinforcement Learning (RL) to improve LLMs' reasoning capabilities. The intermediate outputs make LLMs' decision process more transparent and result in more reliable conclusions \cite{chu2025qwen}. Inspired by these models' success in complex reasoning, we propose MORE-R1, a two-stage training model to guide LVLM to complete the MORE task effectively. MORE-R1 leverages the advanced Qwen2.5-VL \cite{qwen2.5vl} model for general multimodal understanding as the LVLM backbone. We first employ Supervised Fine-Tuning (SFT) for cold-start training, enabling the LVLM to acquire a preliminary stepwise reasoning pattern tailored to the MORE task. Subsequently, we apply RL to optimize the LVLM's reasoning trajectory, thereby enhancing its overall reasoning capability.

At the initial cold-start training stage, we select a small subset of the training data and design an approach to automatically generate high-quality reasoning data. Specifically, we reformulate the MORE task as a stepwise progressive reasoning process. By crafting fine-grained guiding instructions, we leverage an expert model \cite{hurst2024gpt} to generate reasoning demonstrations. Through SFT on the generated data, we guide LVLM to learn the fundamental reasoning paradigm for the MORE task. During the subsequent RL stage, we adopt the advanced Group Relative Policy Optimization (GRPO) algorithm \cite{shao2024deepseekmath} for efficient
RL training. To stabilize training and improve reasoning ability on complex cases, we further design a curriculum-like Progressive Sample-Mixing Strategy. As GRPO training progresses, the proportion of hard samples within each mini-batch is gradually increased, guiding the model to incrementally acquire correct reasoning paths for difficult instances. Our main contributions are summarized as follows:
\begin{itemize}
    \item We propose MORE-R1, a generation-based method using LVLM as the backbone with explicit reasoning for the MORE task. MORE-R1 integrates a two-stage SFT\(+\)RL training framework, enhancing model's scalability, reasoning transparency, and complex-scenario competence.
    \item During Stage 1, we design an efficient automatic data construction strategy to generate high-quality stepwise reasoning data tailored to the MORE task for SFT, enabling the model to effectively learn the reasoning paradigm.
    \item We introduce a Progressive Sample-Mixing Strategy during Stage 2, which gradually raises the model's focus on hard samples, effectively stabilizing training and enhancing model's reasoning performance on challenging cases.
    \item To the best of our knowledge, we are the first to effectively adapt LVLM to the MORE task. Comprehensive experiments demonstrate that MORE-R1 achieves State-Of-The-Art (SOTA) results on the MORE benchmark.
\end{itemize}

\section{Related Works}
\subsection{Multimodal Relation Extraction}\label{sec2.1}
Early multimodal RE works \cite{zheng2021multimodal,ifaformer} treated images as auxiliary information while still focusing on extracting relations between textual entities. MORE \cite{moreformer} is a recently proposed task which poses a higher demand on model's cross-modal reasoning ability. Recent advances in the MORE task primarily focus on enhancing fine-grained cross-modal interaction and multimodal fusion within small-scale VLMs, while employing a classification head at the output layer to complete the MORE task. MOREformer \cite{moreformer} enhances multimodal representations by incorporating object attribute features and depth map features extracted from the image. HVFormer \cite{hvformer} leverages mixture-of-experts architecture to effectively integrate multi-granularity cross-modal interaction semantics. CGI-MRE \cite{cgimre} draws inspiration from bioinformatics and employs genetic algorithms to mine cross-modal shared information. MMIB \cite{mmib} leverages the information bottleneck theory to filter modality noise, further enhancing cross-modal fusion. FocalMRE \cite{focalmre} devises a novel focal attention and gating mechanism to enhance image region interactions. REMOTE \cite{remote} proposes a multi-level optimal transport mechanism to effectively fuse multimodal features. Despite their efforts, these methods still face Classification-based Method's challenges introduced in Section \ref{sec1}.

\subsection{Reinforcement Learning for LLMs and LVLMs}
Recent studies \cite{shao2024deepseekmath,chen2025acereason,chu2025qwen} have shown that applying RL for LLMs' and LVLMs' fine-tuning can significantly enhance reasoning abilities for challenging tasks. RL algorithm for these models can be divided into Off-Policy and On-Policy, depending on whether the training samples are generated by current policy model. Off-Policy algorithm such as DPO \cite{rafailov2023direct} decouples the training data from the current-updated policy model by using pre-collected data without the need to generate new samples during RL training, ensuring high sample efficiency. However, the learning process can be unstable and may result in distribution shift. 

On-Policy algorithm couples the training data with the policy model, where the data needs to be generated in real-time by current policy model. This ensures training stability as the learning objective and sampled data come from the same policy model; however, this results in low sample efficiency and high computational cost. PPO \cite{schulman2017proximal} trains both reward and value models to estimate the advantage for each generation and update the policy model within a constrained range to improve training stability. GRPO \cite{shao2024deepseekmath} replaces the value model with an inner-group relative advantages calculation to guide policy model to update, further simplifying the training process and improving computational efficiency. Moreover, recent study \cite{guo2025deepseek} points out that directly applying RL to LLMs without a preliminary cold-start training can lead to unstable outputs, such as producing repetitive, poorly readable, or language-mixed reasoning results.

\section{Methods}
\subsection{Task Formulation and Overall Framework}
Given an input sample \(x\) which contains an image \(I\) with an object \(o\) specified by a bounding box in the image, and a text description \(T\) with an entity \(e\) in the text, the MORE task \cite{moreformer} aims to identify the relation label between the object and the entity \(R = MORE\_Model(I, o, T, e) \). The relation label consists of the entity types of \(o\) and \(e\) with their semantic relation, such as \textit{/per/org/opposed\_to}. Possible entity types are \textit{Person (per)}, \textit{Organization (org)}, \textit{Location (loc)}, and \textit{Miscellaneous (misc)}. By combining different entity types of \(o\) and \(e\), the MORE task includes 21 possible relation types, including type \textit{none}, as shown in Fig. \ref{fig1}.

\begin{figure}[t]
\centering
\includegraphics[width=0.9\textwidth]{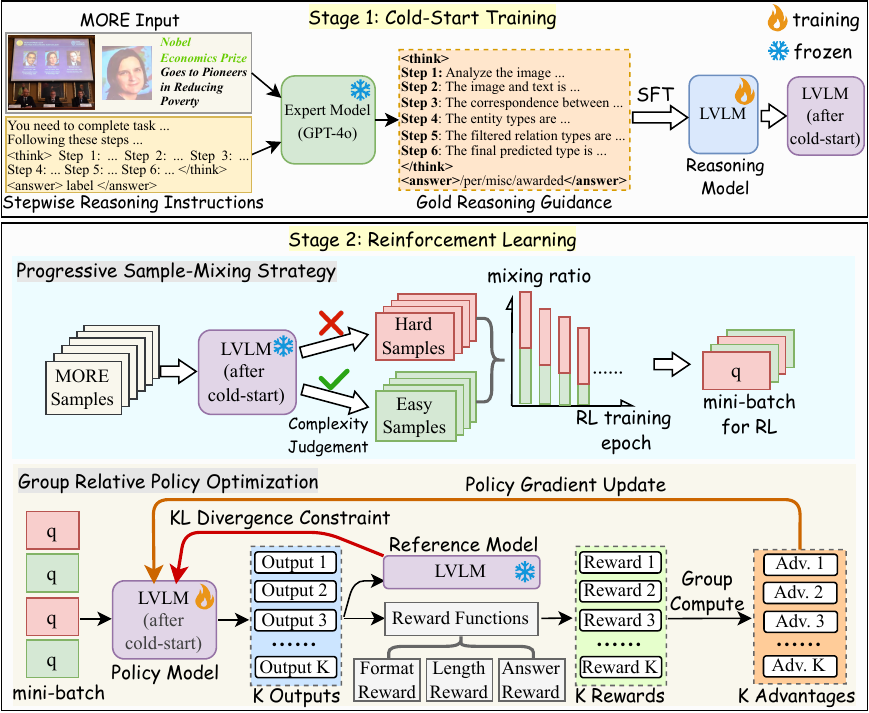}
\caption{Overall framework of MORE-R1, which adopts a two-stage training framework. In Stage 1, the cold-start training enables the LVLM to learn the fundamental reasoning paradigm tailored for the MORE task. In Stage 2, RL further enhances the reasoning capability of the LVLM trained by Stage 1. Models marked with a flame icon indicate that they are involved in training, while the snowflake icon denotes parameter-frozen.} \label{fig2}
\vspace{-1em}
\end{figure}

As shown in Fig. \ref{fig2}, MORE-R1 leverages LVLM as backbone and conducts two-stage training. Stage 1 is cold-start training: we use an expert model to construct a small proportion of high-quality reasoning data, then conduct SFT on LVLM to enable it to acquire basic MORE reasoning ability. Stage 2 is RL: we use the advanced GRPO \cite{shao2024deepseekmath} as the core RL algorithm, and integrate a Progressive Sample-Mixing Strategy. During the RL training process, the ratio of easy and hard samples within each mini-batch is gradually adjusted as the RL training epoch increases. The division of easy and hard samples is determined by the reasoning performance of the LVLM after the Stage 1 cold-start training.

We leverage advanced open-source LVLM Qwen2.5-VL \cite{qwen2.5vl} as the backbone, which has been widely used in recent multimodal research \cite{chu2025qwen,lin2025foundation}. Qwen2.5-VL can accept multiple images with dynamic resolutions as input, using a vision encoder \(\phi_I(\cdot)\) to map visual inputs into textual space, then combined with tokenized text and feeds into the Qwen2.5 LLM \(f(\cdot)\) to generate natural language response. To adapt Qwen2.5-VL to MORE scenario, we crop the object region in the image and input it into the Qwen2.5-VL model alongside the original image as a pair of images. Therefore, employing Qwen2.5-VL as the backbone to perform the MORE task in MORE-R1 can be formulated as:
\begin{equation}
    Rea, Ans=f(\phi_I(I),\phi_I(o),Ins(T,e);\Theta)
\end{equation}
where \(Rea\) and \(Ans\) denote the intermediate reasoning and final predicted label by LVLM, respectively. \(I\) is the original given image, and \(o\) represents the cropped image region corresponding to the object for RE. \(Ins(\cdot)\) is the textual input, which contains task-related instructions with given text \(T\) and entity \(e\) for RE. \(\Theta\) denotes the parameters of the LLM in LVLM.

\subsection{Cold-Start Training Stage}
As shown in Fig. \ref{fig2} upper part, in Stage 1, we guide LVLM to obtain basic stepwise reasoning ability for the MORE task through SFT on small-scale example reasoning data. Since manually annotating reasoning process examples is extremely time-consuming and labor-intensive, we design an efficient automatic reasoning example construction strategy. By crafting fine-grained annotation prompt, we leverage GPT-4o \cite{hurst2024gpt} as expert model, which possesses strong instruction-following ability, to generate reasoning process demonstration data. The annotation prompt for the expert model consists of three parts: basic task description, stepwise reasoning instruction, and actual answer prompt\footnote{The complete prompt template for the expert model is available at https://github.com/MartinYuanNJU/MORE-R1}.

\textbf{Basic task description}: Briefly describes the input modalities and the fundamental aims of the MORE task, lists all candidate relation labels, and instructs the model to output the intermediate reasoning steps in \texttt{<think>} \texttt{</think>} tags, and output the final predicted label in \texttt{<answer>} \texttt{</answer>} tags.

\textbf{Stepwise reasoning instruction}: Contains fine-grained stepwise reasoning instructions with six sequential sub-steps' guidance, enabling the model to incrementally unravel the intricate interdependencies in the MORE input:
\begin{itemize}
    \item \texttt{Step 1}: Image and object analysis. Guide the model to acquire a basic understanding of the content depicted in the image and the roles of the objects within it, serving as the foundation for subsequent reasoning.
    \item \texttt{Step 2}: Cross-modal relevance assessment. Guide the model to assess whether the semantics of the image and the text are correlated, thereby preparing for the subsequent inference of fine-grained correspondences between them.
    \item \texttt{Step 3}: Cross-modal alignment. Guide the model to bridge the correspondences between visual objects and textual entities, which facilitates subsequent reasoning sub-steps. For example, as illustrated in Fig. \ref{fig1}, if the model can recognize that the object in the given image corresponds to the entity ``Celtics'' mentioned in the text, it will be able to accurately infer that the object in image is playing against a team member from ``Heat''.
    \item \texttt{Step 4}: Entity type identification. Guide the model to determine the type of object and entity from \textit{per}, \textit{org}, \textit{loc}, \textit{misc}, enhancing model's understanding of their semantics and further help narrowing down relation candidates.
    \item \texttt{Step 5}: Preliminary relation type filtering. As the relation label contains entity types, this step guides model to narrow down the set of candidate relations, thereby reducing the task complexity. For instance, if the model determines that the object and the entity belong to the \textit{per} and \textit{org} types respectively, then only 4 relations (\textit{opposed to}, \textit{leader of}, \textit{member of}, \textit{none}) out of the 21 candidate labels are possible under the current type combination.
    \item \texttt{Step 6}: Precise relation type determination. Based on the previous steps' multimodal understanding and cross-modal bridging, this step guides model to precisely decide the final relation type from the filtered candidate set.
\end{itemize}

\textbf{Actual answer prompt}: To ensure the expert model arrives at the correct relation through logically coherent intermediate reasoning, we append the ground-truth answer to the end of the overall prompt and require the model to generate reasoning that leads to the label. For non-\textit{none} relations samples, we provide the gold entity types and the relation type. For \textit{none} relation, we prompt the model that there is no relation between the given object and entity.

As GPT-4o is closed-sourced and is accessible only via a paid API\footnote{https://platform.openai.com/docs/models/gpt-4o}, to balance the API usage cost while obtaining a relatively sufficient cold-start training data, we sample 25\% of the samples from each relation category in the training set. After using the expert model to generate reasoning data, we design a simple regex-based filtering strategy to remove samples that do not conform to the reasoning structure or contain incorrect answers. The filtered data is then used for Stage 1 SFT, where the LVLM is provided only with the basic task description. We use the standard next-token prediction loss \cite{chen2024next} for Stage 1 training. 

\subsection{Reinforcement Learning Stage}
Through Stage 1, the LVLM acquires basic reasoning capabilities for the MORE task. To further enhance its reasoning performance in complex scenarios within the MORE task, we conduct RL for Stage 2, which strengthens the model's ability to explore intermediate reasoning paths and discover effective strategies for handling complex cases. Specifically, our RL training adopts the advanced GRPO algorithm \cite{shao2024deepseekmath}, which has been successfully applied to other multimodal understanding tasks such as visual question answering \cite{chu2025qwen} and medical image analysis \cite{lin2025foundation}. During the RL training, we integrate a Progressive Sample-Mixing Strategy to guide the LVLM to effectively master reasoning over complex cases.

\subsubsection{Group Relative Policy Optimization}
GRPO \cite{shao2024deepseekmath} is an on-policy rule-based RL algorithm that simplifies the reward mechanism by removing the need for training value models for reward signal calculation, thereby reducing overall computational complexity. It utilizes direct, rule-based reward functions to evaluate output correctness, guiding models to autonomously explore reasoning paths within the solution space without relying on human-annotated preference data. It possesses significant advantages in tasks with clearly defined correct answers and limited fine-grained annotations \cite{chu2025qwen}. Therefore, GRPO is well-suited for the MORE task, as the final answer is a clearly defined label, and it is extremely challenging and labor-intensive to annotate fine-grained gold reasoning paths.

As shown in the lower part of Fig. \ref{fig2}, GRPO implements a simple but effective computation process: it directly evaluates the relative advantages of multiple possible outputs generated for one input question. Specifically, given an input question \(q\), GRPO first uses old policy \(\pi_{\theta_{\text{old}}}\) to generate \(K\) different responses \(\{o_1, o_2, ..., o_K\}\) (called a group), then calculates reward for each response in current group using a rule-based reward function \(r(\cdot)\), therefore obtains \(K\) corresponding rewards \(\{r(o_1), r(o_2), ..., r(o_K)\}\). Then GRPO calculates the relative advantage score \(A_i\) of each response \(o_i\) through inner-group standardization:
\begin{equation}\label{eqa1}
    A_i = \frac{r(o_i)-\text{mean}(\{r(o_1), r(o_2), ..., r(o_K)\})}{\text{std}(\{r(o_1), r(o_2), ..., r(o_K)\})}
\end{equation}

Based on the relative advantage scores of all responses, GRPO maintains a balance between reward maximization and policy stability by maximizing the following objectives to optimize the current policy model \(\pi_{\theta}\):
\begin{equation}\label{eqa2}
    \mathcal{J}_{\text{GRPO}}(\theta) = \frac{1}{K} \sum_{i=1}^{K} \Big(\text{min} \big(\xi_i A_i, \text{clip}\left(\xi_i, 1-\epsilon, 1+\epsilon\right) A_i \big)- \beta\mathbb{D}_{KL} \big(\pi_{\theta}||\pi_{\text{ref}}\big)\Big)
\end{equation}
where \(\xi_i=\frac{\pi_{\theta}(o_i|q)}{\pi_{\theta_{\text{old}}}(o_i|q)}\) quantifies the policy change, \(\epsilon\) and \(\beta\) control the clipping threshold and the intensity of divergence penalty, respectively. \(\pi_{\text{ref}}\) is the reference policy model to constrain the policy model \(\pi_{\theta}\)'s exploration, avoiding excessive divergence and preventing catastrophic forgetting. Such constraint is enforced via the Kullback–Leibler (KL) divergence term in Equation \ref{eqa2}:
\begin{equation}
    \mathbb{D}_{KL} (\pi_{\theta}||\pi_{\text{ref}}) = \frac{\pi_{\text{ref}}(o_i|q)}{\pi_{\theta}(o_i|q)} - \log \frac{\pi_{\text{ref}}(o_i|q)}{\pi_{\theta}(o_i|q)} - 1
\end{equation}

\textbf{Reward function in GRPO for the MORE task}: The reward function \(r(\cdot)\) for the MORE task is a rule-based function aligning with MORE-R1's learning objective for exploring a reasonable reasoning path to arrive at the correct relation prediction. We design a comprehensive reward which consists of three components: format reward, length reward, and answer reward.
\begin{itemize}
    \item Format reward \(r_{\text{format}}\): This reward aims to enforce the policy model to follow the reasoning template format for completing reasoning exploration. The format is described as: \texttt{<think> Step 1:\([r1]\) Step 2:\([r2]\) Step 3:\([r3]\) Step 4:\([r4]\) Step 5:\([r5]\) Step 6:\([r6]\) </think> <answer>\([label]\)</answer>}, wh-
    ere \([r1]\sim [r6]\) is the unconstrained reasoning content, and the predicted \([label]\) must be one of the 21 candidate relation labels. If the policy model output adheres to such format, the reward is 1.0; otherwise is 0.0.
    \item Length reward \(r_{\text{length}}\): We expect the policy model to think thoroughly at each step to avoid making a hasty judgment, forming a relatively long CoT reasoning for a more accurate final prediction. If the output text length exceeds 1,024, then this reward is 1.0; otherwise is 0.0.
    \item Answer reward \(r_{\text{answer}}\): Since the MORE task has a clear standard answer for relation labels, this reward aims to enable the model to explore the intermediate reasoning path that leads to a correct answer. If the predicted relation label is correct, the reward is 1.0; otherwise is 0.0.
\end{itemize}

Under this composite reward function, LVLM can explore thoroughly under the guidance of the stepwise reasoning paradigm, achieving a more accurate extraction. Consequently, the reward for a given response \(o_i\) is calculated by:
\begin{equation}
    r(o_i) = r_{\text{format}}(o_i)+r_{\text{length}}(o_i)+r_{\text{answer}}(o_i)
\end{equation}

\subsubsection{Progressive Sample-Mixing Strategy}\label{progressives}
In Stage 2, an intuitive way is to use all samples in training set except those used in the Stage 1 SFT for GRPO training (i.e., the remaining 75\% of the original training set). However, when we directly applied these samples to GRPO training with the LVLM after the Stage 1 training, we only observed a slight improvement in model performance, and the answer reward did not show significant growth during GRPO training. Therefore, we proposed a hypothesis: the majority of the remaining 75\% samples are ``easy'', meaning that after Stage 1 training, LVLM can already answer correctly on these samples. Such samples provide little benefit for RL training, and mixing too many in a mini-batch causes model to overfit on easy patterns, impairing its ability to explore complex scenarios. Thus we made further exploration. We used the LVLM after Stage 1 to perform inference on the remaining 75\% of samples and divided the original GRPO training data into easy and hard samples based on the predicted label's correctness. The results showed that 79\% of the original GRPO training samples were easy samples, aligning with our hypothesis. Then a straightforward approach would be training GRPO solely on the hard samples. However, we observed that as training progressed, the answer reward showed almost no improvement, and the model's performance even deteriorated. We further analyzed that without the warm-up guidance of easy samples at the beginning of training, the model failed to discover effective exploration trajectories for complex cases, leading to ineffective policy updates.

\begin{algorithm}[t]
  \small
  \caption{GRPO with Progressive Sample-Mixing Strategy}
  \textbf{Input} LVLM after cold-start training $\pi_{\theta_{\text{init}}}$; original GRPO training set $\mathcal{D}$; reward function \(r(\cdot)\); hyperparameters $\epsilon$, $\beta$, $\mu$, $\alpha$; mini-batch size $B$
  \begin{algorithmic}[1]
    \State using $\pi_{\theta_{\text{init}}}$ to divide $\mathcal{D}$ into easy samples set $\mathcal{D_{\text{easy}}}$ and hard samples set $\mathcal{D_{\text{hard}}}$
    \State reference model $\pi_{\text{ref}} \leftarrow \pi_{\theta_{\text{init}}}$
    \State policy model $\pi_\theta \leftarrow \pi_{\theta_{\text{init}}}$
    \For{epoch \(t\) = \(1, 2, \dots, E\)}
      \For{step = \(1, 2, \dots, M_t\)}
      \State sample \(\lceil \frac{\alpha^{t-1}B}{1+\alpha^{t-1}} \rceil\) data from $\mathcal{D_{\text{easy}}}$, sample \(\lceil \frac{B}{1+\alpha^{t-1}} \rceil\) data from $\mathcal{D_{\text{hard}}}$
      \State mixing the sampled data to form a mini-batch $\mathcal{D}_b$ 
      \State update the old policy model $\pi_{\theta_{\text{old}}} \leftarrow \pi_{\theta}$ 
      \State sample $K$ outputs $\{o_i\}_{i=1}^K \sim \pi_{\theta_{\text{old}}} (\cdot \mid q) $ for each question $q \in \mathcal{D}_b$
      \State compute rewards $\{r_i\}_{i=1}^{K}$ for each sampled output $o_i$ using \(r(\cdot)\) 
      \State compute relative advantage score $\{A_i\}_{i=1}^{K}$ for each output $\{o_i\}_{i=1}^{K}$
      \For{GRPO iteration = \(1, 2, \dots, \mu\)}
        \State update the policy model $\pi_{\theta}$ by maximizing the objective \(\mathcal{J}_{\text{GRPO}}(\theta)\)
      \EndFor
    \EndFor 
    \EndFor 
  \end{algorithmic}
  \textbf{Output} LVLM policy model after GRPO training $\pi_\theta$
  \label{alg:iter-grpo}
\end{algorithm}

Based on the observations above, we propose a Progressive Sample-Mixing Strategy during the RL training process. As shown in Fig. \ref{fig2} middle part, we first conduct complexity judgement by the LVLM after Stage 1 training, dividing original GRPO training samples into hard and easy ones, then during GRPO training, the ratio of easy and hard samples within each mini-batch is gradually adjusted as the training epoch increases. Specifically, the proportion of hard samples is progressively increased, enabling the model to transition smoothly toward learning from more challenging examples and thereby effectively enhancing its reasoning ability. At the initial GRPO training epoch, we set the ratio of easy to hard samples in each mini-batch as 1:1. As the number of training epoch \(t\) increases, the ratio of easy samples gradually decreases, formulated as:
\begin{equation}\label{eqamine}
    ratio_{\text{easy}}(t):ratio_{\text{hard}}(t) = \alpha^{t-1}:1
\end{equation}
where \(t \geq 1\) is the current training epoch, \(\alpha \in (0,1]\) is the decay factor. We illustrate the overall computation logic of Stage 2 in Algorithm \ref{alg:iter-grpo}.

\section{Experiments}
\subsection{Experimental Settings}
\textbf{Dataset and Evaluation}: We conduct evaluation on the commonly-used MORE benchmark \cite{moreformer}. The MORE dataset used online multimodal news data as resources and annotated 20,264 samples. Each sample contains an image and a text, with a given object (by bounding box) in image and a given entity in text, and the gold answer for the relation between the object and the entity. The MORE benchmark provides 21 possible relation types (including \textit{none} type), and splits all samples into 15,486 for training, 1,742 for development, and 3,036 for test. Following previous works \cite{ifaformer,moreformer,hvformer,focalmre,cgimre,mmib,remote}, we report Accuracy (\(Acc\)), Precision (\(P\)), Recall (\(R\)), and F1 Score (\(F1\)) for performance evaluation. As stated by \cite{remote}, because MORE is a class-imbalanced task, \(F1\) is the most important metric as it relieves the long-tail effect and provides a robust evaluation.

\textbf{Baselines}: We compare MORE-R1 with current SOTA methods for MORE task, including IFAformer \cite{ifaformer}, MOREformer \cite{moreformer}, HVFormer \cite{hvformer}, MMIB \cite{mmib}, CGI-MRE \cite{cgimre}, FocalMRE \cite{focalmre}, REMOTE \cite{remote}. All these works are Classification-based and details of these works have been introduced in Section \ref{sec2.1}. We design an LVLM baseline Qwen2.5-VL-SFT by directly using the MORE training set to SFT the same LVLM backbone used by MORE-R1, forcing the model to directly output relation labels, which belongs to the Generation-based (no reasoning) Method. We also report representative LVLMs' zero-shot inference results by directly prompting LVLM to output the relation label without SFT.

\textbf{Implementation Details}: We use ms-swift 3.8.0 framework for overall training, and all training is conducted on 8 NVIDIA A100 80GB GPUs. We use the widely adopted Qwen2.5-VL-7B-Instruct version as our LVLM backbone. In all training, we only fine-tune the LLM part in LVLM and freeze the vision encoder and projection layer. For Stage 1 training, we sampled 25\% samples from each relation category in training set, and used GPT-4o to create reasoning annotations. After filtering out responses that did not meet the output requirements, we obtained 3,865 samples for training. In Stage 1, MORE-R1 is trained with 20 epochs, per GPU batch size 8, learning rate 1e-4, gradient accumulation steps 2, warmup ratio 0.05, max length 2096. Stage 1 training costs about 2 hours.

We use the LVLM after Stage 1 training to infer on the remaining training set to divide the samples into 2,370 hard ones and 8,853 easy ones based on the predicted label correctness. During Stage 2, we set 4 epochs for training, and set the decay factor \(\alpha\) for progressive sample-mixing as 0.5. In each GRPO training epoch, the data consist of all hard samples mixed with a number of easy samples determined by the ratio defined by Equation \ref{eqamine}. When sampling easy samples, we ensure that the proportion of \textit{none} and non-\textit{none} category samples matches that of the original training set. The GRPO is trained on LVLM after Stage 1 training with per GPU batch size 16, learning rate 1e-6, gradient accumulation steps 1, warmup ratio 0.05, temperature 0.8, max completion length 1024. \(\epsilon,\beta,\mu\) in GRPO Algorithm \ref{alg:iter-grpo} are set as 0.2, 0.001, 2, respectively. Stage 2 training costs about 24 hours. All stages are trained with bfloat16 precision. For inference, we use vLLM engine for acceleration and set temperature to 0 for stable and reproducible outputs. Inference on the entire test set takes approximately 7 minutes on a single A100 GPU.

\subsection{Main Results}

\begin{table}[t]
\centering
\caption{Results of MORE-R1 and its variants and other SOTA baselines on the MORE benchmark. The best result is \textbf{bolded} and the second best is \underline{underlined}. ``(-zs)'' denotes zero-shot results without training. ``Qwen2.5-VL-SFT(25\%)'' denotes using the same 25\% data in training set for Stage 1 to conduct label SFT without reasoning training. ``MORE-R1(Stage 1)'' retains only cold-start training without Stage 2 RL.}
\begin{tabular}{l|cccc}
\toprule
\textbf{Model} & \textbf{  Accuracy } & \textbf{Precision} & \textbf{   Recall   } & \textbf{F1 Score } \\ 
\hline
\multicolumn{5}{c}{Classification-based Method} \\
\hline
 IFAformer \cite{ifaformer} & 79.28 & 55.13 & 54.24 & 54.68 \\
 MOREformer \cite{moreformer} & 83.50 & 62.18 & 63.34 & 62.75 \\
 HVFormer \cite{hvformer} & 82.31 & 58.81 & 62.84 & 60.76 \\
 MMIB \cite{mmib} & 82.41 & 60.15 & 62.28 & 61.17 \\
 CGI-MRE \cite{cgimre} & 81.92 & 57.44 & 63.01 & 60.09 \\
 FocalMRE \cite{focalmre} & 82.44 & 60.75 & 62.89 & 61.81 \\
 REMOTE \cite{remote} & \underline{83.64} & \underline{63.21} & 64.63 & \underline{63.91} \\
\hline
\multicolumn{5}{c}{Generation-based (no reasoning) Method} \\
\hline
 LLaMA3.2-Vision-11B(-zs) \cite{llama3.2vision} & 6.69 & 5.09 & 18.66 & 8.01 \\
 Qwen2.5-VL-7B(-zs) \cite{qwen2.5vl} & 23.94 & 25.81 & 23.95 & 24.84 \\
 Qwen2.5-VL-SFT & 82.64 & 55.08 & \underline{64.84} & 59.56 \\
 Qwen2.5-VL-SFT(25\%) & 80.20 & 51.85 & 54.11 & 52.96 \\
\hline
\multicolumn{5}{c}{Generation-based (with reasoning) Method} \\
\hline
 MORE-R1(Stage 1) & 83.33 & 62.53 & 62.84 & 62.69 \\
 MORE-R1 & \textbf{84.91} & \textbf{65.88} & \textbf{69.83} & \textbf{67.80} \\
\bottomrule
\end{tabular}
\label{tab1}
\vspace{-1em}
\end{table}

Table \ref{tab1} presents the performance of our MORE-R1 and current SOTA baselines on the MORE benchmark. The results show that our method significantly outperforms current SOTA baselines. Compared with SOTA classification-based method REMOTE \cite{remote}, the complete MORE-R1 outperforms it by 1.5\%, 4.2\%, 8.0\%, 6.1\% on \(Acc, P, R, F1\) respectively. Compared with generation-based (no reasoning) baseline Qwen2.5-VL-SFT, MORE-R1 surpasses it by 2.7\%, 19.6\%, 7.7\%, 13.8\% on \(Acc, P, R, F1\). These results prove the effectiveness of MORE-R1 to perform explicit reasoning before generating the final extraction answer and use a two-stage training paradigm to enhance LVLM's reasoning ability for the MORE task. Moreover, we observe that (1) Qwen2.5-VL-SFT, the baseline which uses all training set to directly train LVLM to output relation label, is significantly inferior to current SOTA model REMOTE, with a 12.9\% lower \(P\) and a 6.8\% lower \(F1\); (2) LVLMs perform very poorly under zero-shot prompting scenario (``(-zs)'' in the Table, the results are reported by \cite{remote}). These indicate that LVLM cannot achieve ideal results with simple prompt learning or direct fine-tuning on MORE task, which further highlights the advancement of our work for successfully adapting LVLM to MORE task with SOTA performances.

\subsection{Ablation Studies}
We conduct several ablation studies to validate the effectiveness of each component in our method. For Stage 1 cold-start training, we report its performance in Table \ref{tab1} as ``MORE-R1(Stage 1)''. We can observe that after Stage 1 training, MORE-R1 is relatively close to SOTA results REMOTE \cite{remote}, with \(Acc\) and \(F1\) being only 0.4\% and 1.9\% lower. Compared with Qwen2.5-VL-SFT, MORE-R1(Stage 1) surpasses it by 0.8\% and 5.3\% on \(Acc\) and \(F1\), and this is achieved with only 25\% of training data compared with REMOTE and Qwen2.5-VL-SFT. These indicate that LVLM acquires basic reasoning ability for the MORE task through SFT. With further Stage 2 RL training, MORE-R1 gains a significant boost in reasoning ability, outperforming all baselines by a large margin.

Moreover, we also observe that compared with Qwen2.5-VL-SFT, MORE-R1(Stage 1) significantly outperforms it in \(P\) by 13.5\% higher while exhibiting a slight drop in \(R\) by 3.1\% lower. We explain this as follows. As Qwen2.5-VL-SFT directly SFT all data in training set only using the final relation label as supervised signal, the model failed to effectively learn the semantic gap between \textit{none} category and the other 20 non-\textit{none} categories which contain actual relation semantics, treating all 21 categories as discrete and equally weighted labels. Thus the model tends to predict a non-\textit{none} relation for those complex samples which actually contain no relation. Overall, this leads the model to favor predicting one of the 20 non-\textit{none} relations, which indirectly increases \(R\) but significantly decreases \(P\). Using only 25\% samples compared with Qwen2.5-VL-SFT, MORE-R1(Stage 1) can achieve a more robust discrimination between \textit{none} and non-\textit{none} relations through stepwise reasoning guiding, achieving a 5.3\% higher \(F1\) for overall evaluation. Furthermore, we report Qwen2.5-VL-SFT(25\%) using the same 25\% samples in Stage 1 to conduct label-only SFT for a fairer comparison, and find that MORE-R1(Stage 1) significantly outperforms it on all metrics.

\begin{figure}[t]
\centering
\includegraphics[width=1.0\textwidth]{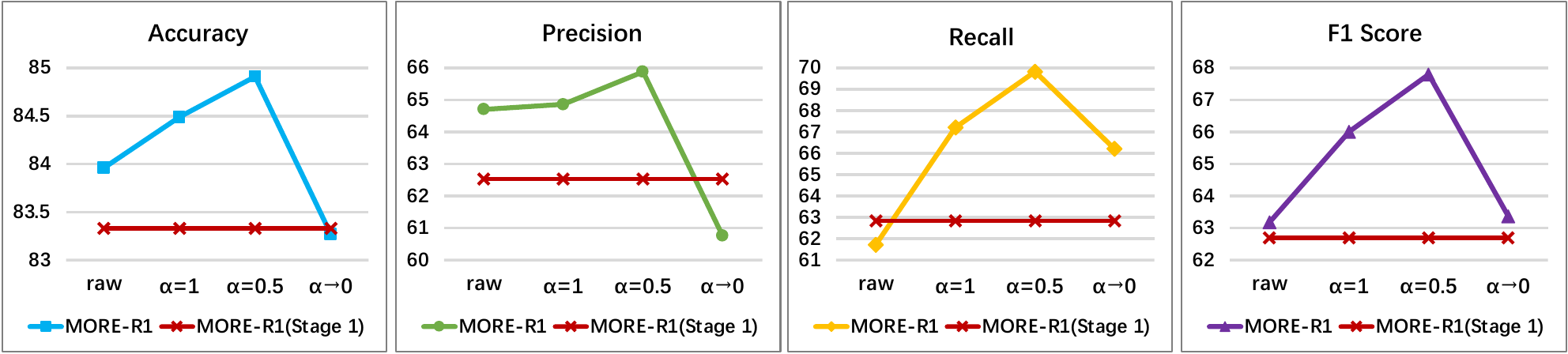}
\caption{Results of different sample-mixing strategies on four evaluation metrics. ``MORE-R1(Stage 1)'' is the baseline that only conducts Stage 1 training. ``MORE-R1'' reports the result after Stage 2 training with different strategies.} \label{fig4}
\vspace{-1em}
\end{figure}

To validate the usefulness of our Progressive Sample-Mixing Strategy in Stage 2, we design three variants: (1) ``raw'': directly using the remaining 75\% of the original data after Stage 1 for each GRPO training epoch, mixing all the easy samples with hard samples in every epoch. (2) ``\(\alpha=1\)'': sampling an equal amount of easy samples as hard samples in each training epoch, ensuring a 1:1 mixing ratio without progressive decay. (3) ``\(\alpha \to 0\)'': only training hard samples in each epoch. As shown in Fig. \ref{fig4}, in each evaluation metric, our complete MORE-R1 which sets decay factor \(\alpha=0.5\) for progressive easy-sample decreasing achieves the best among all variants. The main reason for the suboptimal performance of ``raw'' and ``\(\alpha\to 0\)'' has been analyzed in Section \ref{progressives}. We also observe that for ``\(\alpha\to 0\)'', the \(Acc\) and \(P\) performances are even lower than that before RL training, and for ``raw'', the \(R\) metric is lower than MORE-R1(Stage 1). We explain these phenomena as follows. The easy samples contain a large amount of \textit{none} type data (7,417 of 8,853), while most of the hard samples are non-\textit{none} (1,574 of 2,370). During RL, mixing all the easy samples with hard ones (``raw'') causes the model to optimize on numerous less informative \textit{none} samples, thereby weakening its reasoning ability for non-\textit{none} cases and leading to a decrease in \(R\). Similarly, only retaining hard samples for RL (``\(\alpha\to 0\)'') severely weakens its ability to distinguish between \textit{none} and non-\textit{none} cases, thereby lowering its \(Acc\) and \(P\). The superior results of our original Progressive Sample-Mixing Strategy (\(\alpha=0.5\)) prove its effectiveness, as it helps LVLM form stable exploratory behaviors and progressively adapt to more complex MORE cases.

\subsection{Case Studies}
\begin{figure}[t]
\centering
\includegraphics[width=1.0\textwidth]{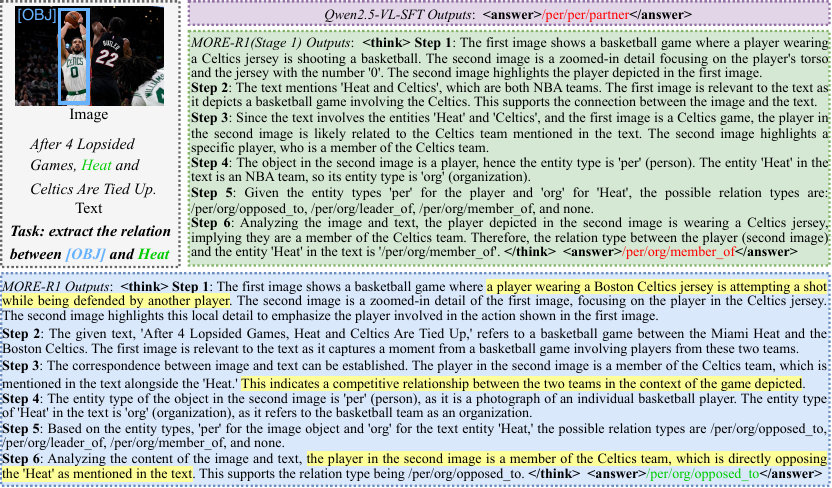}
\caption{Example of actual outputs for Qwen2.5-VL-SFT (purple box), MORE-R1(Stage 1) (green box), and MORE-R1 (blue box). Only MORE-R1 provides the correct answer.} \label{fig5}
\vspace{-1em}
\end{figure}
As shown in Fig. \ref{fig5}, we present an example from the test set and use three models for inference: Qwen2.5-VL-SFT, MORE-R1(Stage 1), and our complete MORE-R1. We can observe that Qwen2.5-VL-SFT predicts incorrectly, and even makes errors in entity types' judgment, indicating its inability to handle complex scenarios. MORE-R1(Stage 1) possesses basic stepwise reasoning ability: it correctly identifies entity types and filters potential candidate relations. However, due to the lack of further RL for reasoning enhancement, the model struggles to infer the implicit adversarial semantics between the cross-modal entities. After progressive sample-mixing RL training, the complete MORE-R1 demonstrates strong reasoning ability in complex scenarios. As depicted in the yellow-highlighted part of Fig. \ref{fig5}, MORE-R1 can uncover the adversarial relationships between entities and consequently identify the correct ``opposed to'' relation. The visualization results further confirm the effectiveness of MORE-R1 for handling MORE task.

\section{Conclusion}
We propose MORE-R1, a generation-based method with two-stage training that adopts LVLM as backbone and conducts explicit reasoning to solve the MORE task. In MORE-R1, we first perform cold-start training by leveraging automatically constructed, fine-grained stepwise reasoning data to SFT the LVLM backbone. This process instills a reasoning pattern that is specifically tailored to the MORE task. Then we perform RL training based on GRPO algorithm to further enhance LVLM's task-specific reasoning ability. During RL, we devise a Progressive Sample-Mixing Strategy to stabilize reasoning optimization and strengthen model's reasoning ability on challenging cases. Extensive experiments on the MORE benchmark demonstrate the effectiveness and superiority of our method. 

\bibliographystyle{splncs04}
\bibliography{ref}
\end{document}